# Stacked Generalization Approach to Improve Prediction of Molecular Atomization Energies

*Ruobing Wang\**

Department of Chemistry and Social Science Research Institute, Duke University, Durham, North Carolina 27708, United States.

AUTHOR INFORMATION

**Corresponding Author**

*ruobing.wang@duke.edu

**Abstract**

Machine learning holds the promise of learning the energy functional via examples, bypassing the need to solve complicated quantum-chemical equations and realizing efficient computing of molecular electronic properties. However, a single machine learning model may reach an upper limit of prediction accuracy even with optimal parameters. Unlike the "bagging" and "boosting" approaches which can only combine the machine learning algorithms of a same type, the stacking approach can combine several distinct types of algorithms through a meta-machine learning model to maximize the generalization accuracy. Here we present a stacked generalization approach for predicting ground state molecular atomization energies. The results suggest prediction error from stacked generalization frameworks are significantly reduced by 38%, compared to the best level-0 individual algorithm that construct the stacked generalization framework. Furthermore, compared to conventional stacked generalization framework, it shows rescanning the original input space by level-1 meta model can extract more information, which can further improve the prediction accuracy.

**TOC GRAPHICS**

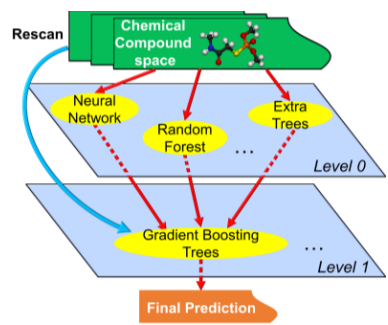

The development of organic molecule-based devices applicable to light emission,[1-3] light-energy conversion,[4] and microelectronics[5] has received considerable attention over the past decades. Paralleling the development of molecular electronic applications, modern electronic structure theories have progressed to the point where the electronic properties of molecules can be routinely calculated with satisfactory accuracy. For instance, the Kohn–Sham scheme of density functional theory has become a popular method to solve electronic structure problems in a wide variety of scientific fields.[6-7] However, the accurate and reliable prediction of properties of molecules typically requires computationally intensive quantum-chemical calculations. In this regard, machine learning holds the promise of learning the energy functional via examples, bypassing the need to solve the Kohn–Sham equations and realizing efficient computing.[8-13]

The atomization energy is an important molecular property that determines the stability of a molecule with respect to the atoms that compose it. Atomization energies are also measurable experimentally and are frequently used to assess the performance of approximate computational methods.[11] Recently, machine learning approach has been demonstrated to predict atomization energies of various small molecules in their given ground-state geometry,[10-11, 14-15] but these studies are using a single machine learning model to maximize the accuracy of predicting atomization energies. In a mathematical point of view, the machine learning algorithm is a sophisticated fit to a non-linear function, and a single machine learning model may fit well to a certain subset of the chemical compound space (CCS), but may overfit or underfit to the rest of the CCS. As a result, the prediction accuracy of a single model may reach an upper limit even with optimal parameters. People are seeking various methods to improve the prediction accuracy for chemical properties, and there are two major approaches: one is to explore different molecular representations[11] or descriptors for CSS[16-17]; the other is to investigate advanced algorithms such as deep learning (neural networks),[11, 18-19] tree-based models[14] and support vector machines.[20] Besides these two approaches, an alternative method is to combine the advantages of several models to break through the upper limit of a single machine learning algorithm (i.e. ensemble method).

Stacked generalization is an ensemble method of using a higher-level model to combine lower-level models to achieve higher predictive accuracy. Unlike the "bagging" and "boosting" approaches which can only combine the machine learning algorithms of a same type, stacked generalization can combine different types of algorithms through a meta-machine learning model to maximize the generalization accuracy. Although most of its current application is to solve classification problems,[21] a stacked generalization framework can function as a regression model as well.[22] Here we present an unprecedented strategy of predicting ground state atomization energies of molecules utilizing a stacked generalization framework, which combines the advantages of distinct algorithms and significantly improves the prediction accuracy by 38%. Furthermore, we improve the strategy of conventional stacked generalization by the method of rescanning original input space, and exploit the general rules of building an effective stacked generalization framework.

The molecular database used in our study has been described in previous literature,[14] which is generated from the PubChem Substance and Compound database.[23-24] The dataset contains 16, 242 molecules and their DFT calculated ground state energies,[14] in which: (i) Each molecule has to be composed of a subset of the elements from the set C, H, N, O, P and S (CHNOPS molecules). (ii) Each molecule must have at least 2 and at most 50 atoms. (iii) The maximum distance between two atoms in a molecule must not exceed 25 $a_0$ ($a_0$= 0.529 Å, i.e., Bohr radius), for the convergence of planewave calculations, where each molecule is placed in a cubic box of side length 30 $a_0$. (iv) There must be an even number of electrons in the molecule. The mean absolute value of ground state atomization energy is 3506.37 kcal/mol. The total variability in the dataset, quantified by the standard deviation of ground state atomization energies, is 1147.89 kcal/mol,[14] which is larger than the variability reported in some of the earlier works in the literature (less than 250 kcal/mol),[11, 15] indicating that a much wider range of molecular systems being included in this study. The distribution of the ground state energy is shown in **Figure S1**. (Supporting information)

In this study, the feature space used to describe the input space consist of two parts. The first part is constructed from the Coulomb matrix,[10] which are defined as

$$C_{IJ} = \begin{cases} 0.5 Z_I^{2.4} & I = J \\ \dfrac{Z_I Z_J}{|R_I - R_J|} & I \neq J \end{cases} \quad (1)$$

where $Z_I$ are atomic numbers, $R_I$ are atomic coordinates, and indices $I, J$ run over the atoms in a given molecule. The Coulomb matrices represent the full set of parameters that DFT calculations take as inputs. Therefore, the problem of predicting atomization energies using $C_{IJ}$ is well defined.[10] The maximum number of atoms for molecules in the dataset is limited to 50, thus the size of Coulomb matrices in this study is 50*50. Molecules with less than 50 atoms have their Coulomb matrices appended by columns and rows of 0 to complete them to have dimensions of 50*50 (**Figure 1a**). To find a unique ordering of the atoms in the Coulomb matrix, the rows (and columns) $C_I$ of the Coulomb matrix are ordered by their norm, i.e. $\|C_I\| \geq \|C_{I+1}\|$, which ensures a unique Coulomb matrix representation (**Figure 1b**).[11] Since the Coulomb matrix is symmetric, the upper triangular part of the matrix is unrolled into a 1275-dimensional vector, which defines the first part of the feature space (**Figure 1c**). [11, 14]

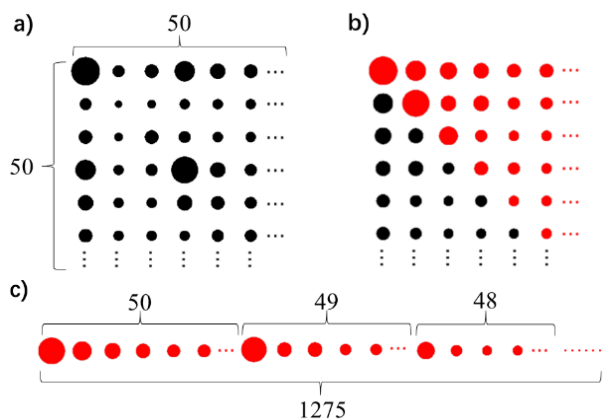

**Figure 1. General steps to generate feature vectors from Coulomb matrix: (a)Original Columbic matrix. (b) Sorted Coulomb matrix. (c) Generated feature vector from Coulomb matrix**

Besides the features derived from Coulomb matrix, the second part of the feature space includes some molecular structural information: molecular weight, number of CHNOPS atoms (six features), and the ratio of carbon atom number vs. sum of other atoms' number. In total, the feature space that used to describe the molecule has a dimension of 1283.

Predicting molecular atomization energies is a regression task, which requires a function or model $f$ that maps an input vector x $\in$ $\mathbb{R}^d$ (feature vectors, d=1283 in this study) onto the corresponding continuous label value y $\in$ $\mathbb{R}$ (here the atomization energy). Since a training data set $\{(x_1, y_1), ..., (x_n, y_n)\}$ is utilized to find $f$, the task falls into the category of a supervised learning problem. The machine learning problem is formulated as a minimization problem of the form:

$$\min_{f} \sum_{i=1}^{n} l(f(x_i), y_i) + \lambda r(f) \qquad (2)$$

The first term of the objective is the empirical risk described by a loss function S which measures the quality of the function f. A specific case is the squared loss $l(\hat{y}, y) = (\hat{y} - y)^2$. The second term of the objective in eq. (2) is a regularization term which measures the complexity or roughness of the function f, which usually is a norm of $f$ or its derivatives. $L_2$ regularization is used in this study.

In this study, 80% of the data is randomly selected as the training set (12993 samples), which is used to train the machine learning model. The rest 20% of the data is utilized as the hold-out test set. The distributions of atomization energies in the training set and test set are displayed in **Figure S2** to ensure the randomness. The model is then built on the training set, and the average loss over the hold-out test set (test error) is utilized as the evaluation for the model performance. The training and evaluation of a single machine learning algorithm has been described in detail by previous literatures.[11, 14] To be consistent with previous studies, the prediction accuracy is evaluated by root mean squared error (RMSE) over the hold-out test set.

**Table 1**. Test RMSE in kcal/mol of single models

| Model | Test RMSE (kcal/mol) |
|---|---|
| NN | 32.34 |
| RR | 35.81 |
| RF | 37.19 |
| ET | 34.71 |
| GBT | 30.74 |

Before establishing stacked generalization framework to ensemble different models, we first test the prediction performance of several single algorithms. Five different machine learning algorithms are trained using the training set and predict on the test set: neural networks (NN), ridge regression (RR), random forest (RF), extremely randomized trees (ET), and gradient boosting trees (GBT). The performance of each model measured by test error (RMSE) is shown in **Table 1**. Although we will not discuss in depth in this article, the result shows the prediction accuracy of NN and GBT are improved by including molecular structure information as features, compared to previous study that only use columbic matrices to construct feature space (The RMSE of NN and GBT on the same data set are 41.81 kcal/mol and 36.63 kcal/mol in previous study,[14] respectively). The implementation details of the single algorithms and corresponding key parameters are described in Supporting Information.

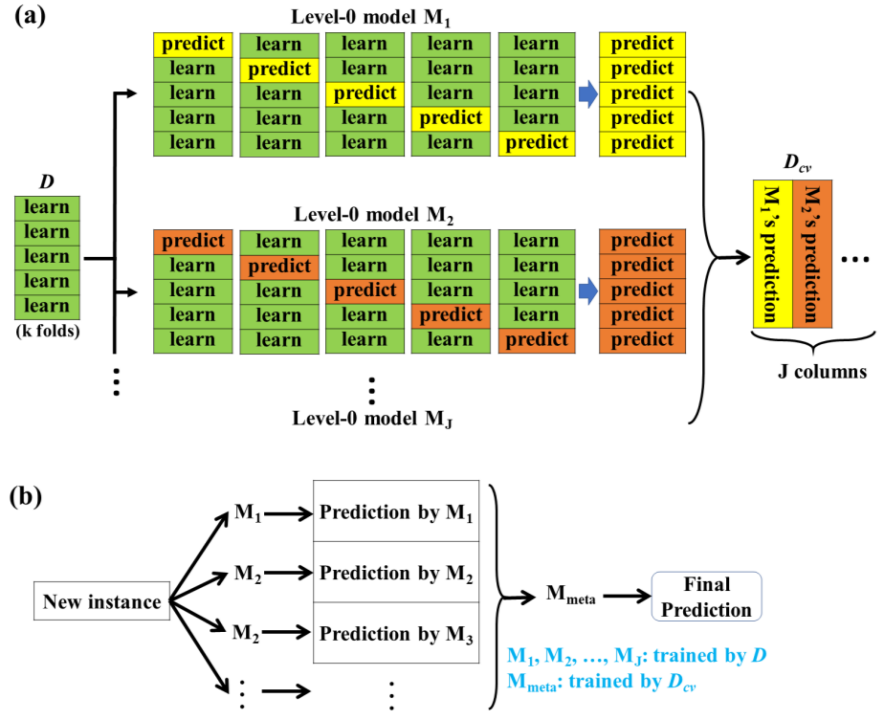

**Figure 2.** (a) K-fold cross validation process to obtain training data ($D_{cv}$) for $M_{meta}$. (b) Prediction process of stacked generalization framework

Stacked generalization is utilized to ensemble different machine learning algorithms, which can be viewed as a means of collectively using several models to estimate their own generalizing biases with respect to a particular learning set, and then filter out those biases.[25-26] There are two kinds of models in a stacked generalization framework: several base models (level-0 models) and one meta-model (level-1 model). The essence of stacked generalization is to use the level-1 model to learn from the predictions of level-0 models. Generally, a stacked generalization framework can obtain more accurate prediction compared to the best level-0 model.[25]

One of the key points is to obtain the training data for level-1 model ($D_{cv}$) from cross-validation technique. The procedures of generating $D_{cv}$ are shown in **Figure 2a**. Given an original data set $D$ = {($y_n$, $x_n$), $n$ = 1, ..., N}, where $y_n$ is the target value and $x_n$ represents feature vectors of the $n$th instance, randomly split the data into $K$ almost equal folds $D_1$, $D_2$, …, $D_K$ ($K$=5 in this study). Define $D_k$ and $D^{(-k)}$ = $D$ – $D_k$ to be the test and training sets for the $k_{th}$ fold of a $K$-fold cross-validation. Given J different

level-0 machine learning algorithms ($M_1$, $M_2$, ..., $M_J$), each $M_j$ is trained by $D^{(-k)}$ and predict each instance $x$ in $D_k$. Let $v_k^{(-j)}(x)$ donate the prediction of the model $M_j$ on $x$. Then we have:

$$z_{kn} = v_k^{(-j)}(x_n) \quad (3)$$

At the end of the entire cross-validation process of each $M_j$, the data assembled from the outputs of the J models is

$$D_{cv} = \{(y_n, z_{1n}, ..., z_{Jn}), n = 1, 2, ..., N\}. \quad (4)$$

$D_{cv}$ is the training set of level-1 model $M_{meta}$. To complete the training process, level-0 models $M_j$ (j=1, 2, ..., J) are trained using original dataset $D$, and $M_{meta}$ is trained by $D_{cv}$.

Now we consider the prediction process, which uses the models $M_j$, j= 1, 2, ..., J, in conjunction with $M_{meta}$. Given a new instance, models $M_j$ produce a vector ($z_1$, ..., $z_J$). This vector is input to the level-1 model $M_{meta}$, whose output is the final prediction result for that instance (**Figure 2b**). This completes the stacked generalization method as proposed by Wolpert.[25]

**Table 2.** Configurations of Stack1, Stack2 and Stack3.

| Framwork | Level-0 model1 | Level-0 model2 | Level-1 model |
|---|---|---|---|
| Stack1 | NN | ET | GBT |
| Stack2 | NN | ET | RR |
| Stack3 | RF | ET | RR |

In order to exploit the mechanism of stacked generalization and investigate the correlation between the predicting accuracy and the framework configuration, the models listed in **Table 1** (using exactly the same parameters) were utilized to establish three different stacked generalization frameworks: **Stack1**, **Stack2** and **Stack3**. Although increasing number of models involved in the stacked generalization framework can potentially improve accuracy, here we use the simplest framework as a proof of principle. The frameworks in this study consist of only two level-0 models and one level-1 model (meta model). **Table 2** exhibits the detailed construction of each framework. Furthermore, each framework has two modes: the 'normal mode' and the 'rescan mode', as shown in **Figure 3**. The 'normal mode' is the conventional feedforward stacked generalization (**Figure 3a**) proposed by Wolpert,[25] while the 'rescan mode' (**Figure 3b**) is a novel design to rescan the original input space by feeding the original data to level-1 model together with the predictions generated by level-0 models (i.e. use both $D$ and $D_{cv}$ to train $M_{meta}$ and input $(x, z_1, ..., z_J)$ to $M_{meta}$ for final prediction). The rationale to rescan the original data space is derived from fact that level-0 models may failed to extract some important information of the original input space, and level-1 model can be utilized to recapture them.

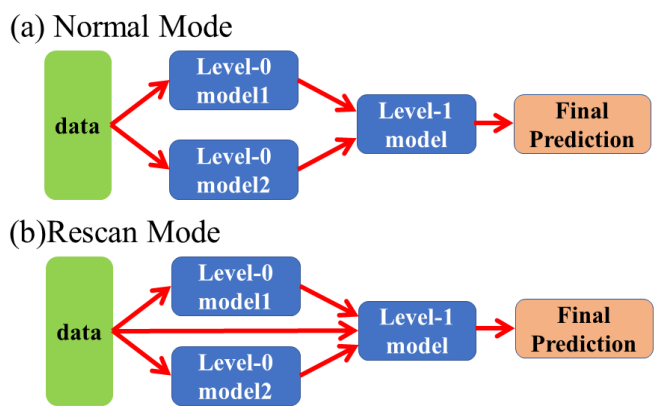

**Figure 3.** General structures of (a) conventional stacked generalization (normal mode) and (b)Stacked generalization with rescanning the original data space by level-1 model (rescan mode)

Using the ensemble training outlined above (Figure 2), we have trained the three stacked generalization frameworks. The test errors, measured by RMSE, for the trained frameworks are summarized in Table 3. The best performance is obtained with **Stack1**'s rescan mode (best RMSE 20.14 kcal/mol), which is improved by 38% compared to its best level-0 model (NN, best RMSE 32.34 kcal/mol), and improved by 45% compared to previous study (best RMSE 36.63 kcal/mol).

**Table 3.** Test RMSE in kcal/mol of stacked generalization frameworks (normal and rescan mode)

| Framework | Mode | Test RMSE (kcal/mol) |
|---|---|---|
| Stack1 | Normal | 27.62 |
|  | Rescan | 20.14 |
| Stack2 | Normal | 28.33 |
|  | Rescan | 28.17 |
| Stack3 | Normal | 33.29 |
|  | Rescan | 29.62 |

**Stack2** and **Stack3** also reduce RMSE compared to their level-0 models. However, the improvement is not significant compared to that of **Stack1**. In order to exploit the difference among **Stack1**, **Stack2** and **Stack3** as well as reveal the general rules to build an effective framework, the hold-out test dataset of 3, 249 molecules are divided to 11 groups based on molecular weights and the molecular weight increases from group #1 to group #11 (i.e. the lowest molecular weight of group #n+1 is larger than the highest molecular weight in group #n). The molecular weight range and number of molecules in each group are listed in **Table S1** (Supporting Information). By tracking the mean absolute errors (MAE) of predictions on each group, the prediction performance curve of a model/framework over different molecular weight range can be calculated. **Figure 4a** exhibits the performance curves of **Stack1** (normal and rescan mode, dash lines) and its level-0 models (NN and ET, solid lines). The performance curves of **Stack2**, **Stack3** and their corresponding level-0 models are also calculated and demonstrated in **Figure 4b** & **4c**, respectively.

In Figure 4a, it can be seen that ET (green solid line) and NN (red solid line) exhibit different predicting performance with respect to different molecular weight ranges. ET has lower MAE when predicting molecules with lower molecular weight (group #1 - #6). However, the accuracy of ET is poor when predicting molecules with higher molecular weights (group #7 - #11), especially for group #9, #10 and #11. NN is defeated by ET when predicting low molecular weight molecules, but its performance is much better than ET in high molecular weight range (group #7 - #11). In other words, ET possesses advantage in predicting atomization energy of CHNOPS molecules within the molecular weight range of 31-168 g/mol, while NN provides better prediction in the molecular weight range of 168-508 g/mol. After NN and ET are ensembled by the level-1 meta model GBT, the performance curve of **Stack1** normal mode (blue dash line, **Figure 4a**) almost overlaps with that of ET when predicting the atomization energy of molecules in group #1-#6, but is closer to the curve of NN for group #7-#11. These results underscore that the stacked generalization approach combines the advantages of ET and NN. Moreover, the performance curve also shows that the MAE is further reduced for all groups when using the 'rescan mode' of **Stack1** (oranges dash line, **Figure 4a**), which indicates that the information lost by ET and NN in level-0 is successfully captured by level-1 model GBT.

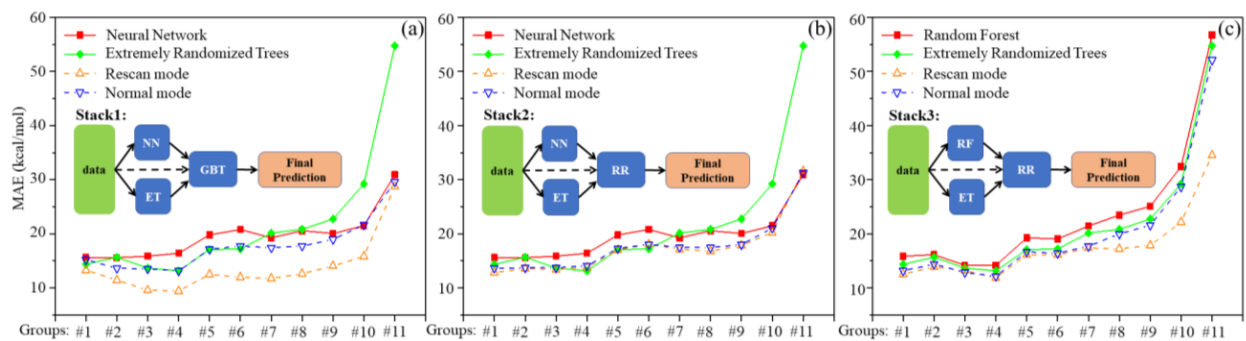

**Figure 4.** Performance curves (based on MAE) of the normal mode and the rescan mode of (a)Stack1, (b)Stack2, (c)Stack3, and the corresponding level-0 models that construct these three stacked generalization frameworks.

The selection of level-0 models can affect the generalization accuracy of stacked generalization. Under the normal mode, **Stack1** and **Stack2** have similar prediction accuracy (RMSE 27.62 kcal/mol and 28.33 kcal/mol, respectively), which is reasonable since the level-1 model only learns from the predictions of level-0 models in the normal mode. In this study, the input vector for level-1 model is in $\mathbb{R}^2$ space (two level-0 models) under the normal mode, and such low dimensional space will not lead to a significant difference between GBT and RR. As a result, the performance curves of **Stack1** and **Stack2** under normal mode are very similar, and both of them combines the advantages of NN and ET. The normal mode of **Stack3** obtains a RMSE of 33.29 kcal/mol, which is not a significant improvement compared to its best level-0 model (ET, RMSE 34.71 kcal/mol), which is due to the similarity between its level-0 models. Both of RF and ET are tree-based models and their performance curves are similar (**Figure 4c**, red and green solid lines). As a result, the normal mode of **Stack3** cannot benefit from the complementary advantages of level-0 models.

Under the rescan mode, unlike **Stack1**, the RMSE of **Stack2** (28.17 kcal/mol) is almost identical to its normal mode (28.33 kcal/mol) and the performance curves overlaps (**Figure 4b**, blue and orange dash lines) which implies the level-1 model RR is unable to extract extra information from the original input space. The possible reason is the linear information that RR can extract from the original dataset is already obtained by NN in level-0. On the other hand, the rescan mode of Stack3 reduce the RMSE compared to its normal model, which indicate RR can extract the linear information that ET and RF failed to obtain in level-0, as shown in **Figure 4c**.

By comparing **Stack1**, **Stack2** and **Stack3**, we can find some general rules of building an effective stacked generalization framework: (i)For level-0 models, it is better to have algorithms with different mechanism (such as tree based model and neural network in this study), which have higher probability of possessing different prediction performance (ii)Rescanning the input space can further improve the prediction accuracy, but the extent of improvement depends on whether the level-1model is complementary to level-0 models.

In conclusion, the prediction accuracy of molecular atomization energy is improved by 38% using ensemble of machine learning models and rescanning the original input space. Without tuning any parameters of single models, the stacked generalization can combine the advantages of level-0 models and rescan the original input space to recapture the lost information to minimize the prediction error. The general mechanism of the stacked generalization is also investigated, which shows the diversity of algorithms is the crucial point of building an effective framework. It is worth underscoring that the stacked generalization strategy is compatible with any other methods of improving machine learning performance (i.e. improving individual algorithms in the framework will also improve the overall accuracy), and can be extended to predicting other molecular properties. Instead of exploring novel machine learning algorithms, the idea of combining conventional models provide a rich possibility of applications for quantum mechanical simulations in computational chemistry and materials science.

ASSOCIATED CONTENT

AUTHOR INFORMATION

Corresponding Author

ruobing.wang@duke.edu

ACKNOWLEDGMENT

The author gratefully acknowledges Dr. Burak Himmetoglu for sharing datasets on open-source platform Kaggle (www.kaggle.com) and Github (www.github.com)

Supporting Information

The detailed implementation of machine learning algorithms and the division of test dataset.